\documentclass[conference]{IEEEtran}
\IEEEoverridecommandlockouts
\usepackage{cite}
\usepackage{amsmath,amssymb,amsfonts}
\usepackage{algorithmic}
\usepackage{graphicx}
\usepackage{graphics}
\usepackage{textcomp}
\usepackage{xcolor}
\usepackage{multirow}
\usepackage{array}
\newcolumntype{L}[1]{>{\raggedright\let\newline\\\arraybackslash\hspace{0pt}}m{#1}}
\newcolumntype{C}[1]{>{\centering\let\newline\\\arraybackslash\hspace{0pt}}m{#1}}
\newcolumntype{R}[1]{>{\raggedleft\let\newline\\\arraybackslash\hspace{0pt}}m{#1}}
\usepackage[super]{nth}
\def\BibTeX{{\rm B\kern-.05em{\sc i\kern-.025em b}\kern-.08em
    T\kern-.1667em\lower.7ex\hbox{E}\kern-.125emX}}
\begin{document}
\makeatletter 
\newcommand{\linebreakand}{%
  \end{@IEEEauthorhalign}
  \hfill\mbox{}\par
  \mbox{}\hfill\begin{@IEEEauthorhalign}
}
\makeatother 

\title{Towards Personalized Brain-Computer Interface Application Based on Endogenous EEG Paradigms

\footnote{
\thanks{This work was the result of project supported by Korea University - KT (Korea Telecom) R\&D Center, and partially supported the Institute of Information \& Communications Technology Planning \& Evaluation (IITP) grant, funded by the Korea government (MSIT) (No. RS-2019-II190079, Artificial Intelligence Graduate School Program (Korea University))}}
}
\author{\IEEEauthorblockN{Heon-Gyu Kwak}
\IEEEauthorblockA{\textit{Dept. of Artificial Intelligence} \\
\textit{Korea University} \\
Seoul, Republic of Korea \\
hg\_kwak@korea.ac.kr} \\
\and 

\IEEEauthorblockN{Gi-Hwan Shin}
\IEEEauthorblockA{\textit{Dept. of Brain and Cognitive Engineering} \\
\textit{Korea University} \\
Seoul, Republic of Korea \\
gh\_shin@korea.ac.kr}
\and 

\IEEEauthorblockN{Yeon-Woo Choi}
\IEEEauthorblockA{\textit{Dept. of Artificial Intelligence} \\
\textit{Korea University} \\
Seoul, Republic of Korea \\
yw\_choi@korea.ac.kr}
\linebreakand

\IEEEauthorblockN{Dong-Hoon Lee}
\IEEEauthorblockA{\textit{Tech. Innovation Group} \\
\textit{KT R\&D Center} \\
Seoul, Republic of Korea \\
donghoon.yi@kt.com}
\and

\IEEEauthorblockN{Yoo-In Jeon}
\IEEEauthorblockA{\textit{Tech. Innovation Group} \\
\textit{KT R\&D Center} \\
Seoul, Republic of Korea \\
yooin.jeon@kt.com}
\and

\IEEEauthorblockN{Jun-Su Kang}
\IEEEauthorblockA{\textit{Tech. Innovation Group} \\
\textit{KT R\&D Center} \\
Seoul, Republic of Korea \\
jun-su.kang@kt.com}
\and

\IEEEauthorblockN{Seong-Whan Lee}
\IEEEauthorblockA{\textit{Dept. of Artificial Intelligence} \\
\textit{Korea University} \\
Seoul, Republic of Korea \\
sw.lee@korea.ac.kr}
}
\maketitle 

\begin{abstract}
In this paper, we propose a conceptual framework for personalized brain-computer interface (BCI) applications, which can offer an enhanced user experience by customizing services to individual preferences and needs, based on endogenous electroencephalography (EEG) paradigms including motor imagery (MI), speech imagery (SI), and visual imagery. The framework includes two essential components: user identification and intention classification, which enable personalized services by identifying individual users and recognizing their intended actions through EEG signals. We validate the feasibility of our framework using a private EEG dataset collected from eight subjects, employing the ShallowConvNet architecture to decode EEG features. The experimental results demonstrate that user identification achieved an average classification accuracy of 0.995, while intention classification achieved 0.47 accuracy across all paradigms, with MI demonstrating the best performance. These findings indicate that EEG signals can effectively support personalized BCI applications, offering robust identification and reliable intention decoding, especially for MI and SI.
\end{abstract}


\section{INTRODUCTION}
Electroencephalography (EEG) is a neurophysiological electrical signal generated by brain neural activity, reflecting various mental states and intentions of an individual. For several decades, researchers have aimed to understand human mental states and intentions by decoding EEG signals, and recent advances in deep learning have significantly improved the performance of EEG decoding \cite{song2022eeg, lee2023toinet, an2023dual, kim2024towards}.
Accordingly, EEG-based brain-computer interface (BCI) technology, which controls external devices based on a user's mental state or intention, has been widely explored in various fields and has been demonstrating promising potential for real-world applications \cite{soleymani2015analysis, vidyaratne2017real, han2020classification, boonyakitanont2020review, lee2020continuous, perslev2021u, lee2023autonomous, kong2023eeg, han2024eeg}.

However, research on personalized BCI applications, which can provide services based on the user's detailed interests, has yet to be widely explored. Most existing BCI systems are designed to function with generalized models that barely consider individual differences in interests, habits, and lifestyles among users, which limits their convenience and user experience. However, the differences in EEG patterns across each individual highlight an opportunity to leverage this user-specific information to build personalized BCI systems that adapt to the unique characteristics and needs of each user.


To achieve this goal, we propose a conceptual framework for personalized BCI application, which can provide service utilizing the user-specific information, through the user identification and intention classification tasks. Moreover, we validate the feasibility of the proposed framework, by conducting preliminary experiments using private endogenous EEG paradigm dataset.  \\

\begin{figure*}[ht]
    \centering
    \includegraphics[width=17cm]{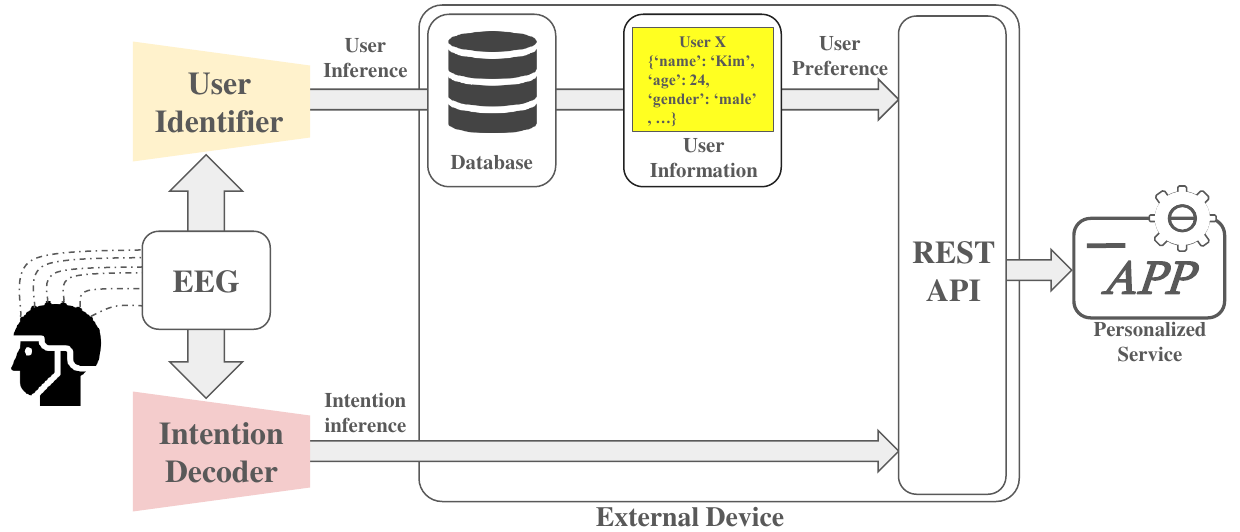}
    \caption{An overview of the proposed framework. The user identification and intention classifier model takes EEG signals as input and returns personalized BCI service based on the user information and intention.}
    \label{fig:fig1}
\end{figure*}

\section{METHODS}
\subsection{Tasks for the Personalized BCI Application}
The main goal of the proposed framework is to provide personalized BCI applications for each user. To achieve this goal, we assumed that the application should have two essential capabilities to perform two main tasks: \textit{i}) user identification to understand user preferences and \textit{ii}) intention classification to execute the user's desired action.
\subsubsection{User identification task}
User identification is the task of determining who is using this application \cite{sun2019eeg}, which can serve as a preliminary step for understanding user preferences. Like other biosignals, EEG is characterized by subject variability \cite{kim2019subject, wu2020transfer, ma2023cross}, meaning that EEG features can vary between individuals even when they are in the same mental state or performing the same task. By leveraging this characteristic, user-specific EEG features can be extracted and utilized for identification, allowing the BCI application to adapt to user-specific features
\subsubsection{Intention classification task}
Intention classification is the task of determining the user's desired or intended action \cite{zhang2019making}. EEG reflects the mental activity of the brain, and therefore can denote the individual's mental state or intention. In particular, endogenous EEG paradigms such as motor imagery (MI), speech imagery (SI), and visual imagery (VI) can provide a more direct and intuitive approach for communicating with BCI applications, compared to exogenous EEG paradigms (e.g., event-related potentials and steady-state visual evoked potentials) \cite{lee2019towards, padfield2019eeg, lee2020neural, palumbo2021motor, lee2021decoding}.

\subsection{Framework for Personalized BCI Application}
Fig 1. illustrates the overall process of the proposed framework. This framework is designed for the BCI application to provide user-intended services in a personalized manner. When a user issues a query to the BCI application through an endogenous EEG paradigm, the system processes the corresponding EEG signals as input. Based on this input, the user identification model retrieves user-specific information from a database and analyzes the user's preferences. Simultaneously, the intention identification model also processes the EEG signals to identify the user's intended action, subsequently calling the appropriate application programming interface (API) associated with the user's intention. By integrating the user's preferences with the identified API, the BCI application can provide a personalized BCI service. This procedure can be represented by the following equations,
\begin{align}
    &p(s) = f(x), \\
    &p(c) = g(x), \\
    &p(c_s) = h(p(p(c) | p(s))),
\end{align}
where $x$ is the input EEG signal, $f(\cdot)$ and $p(s)$ represent an user identification model and the predicted user class, $g(\cdot)$ and $p(c)$ denote an intention classification model and the predicted user class, $h(p(p(c) | p(s)))$ is an API calling process conditioned on the user prediction, and $p(c_s)$ represents the output personalized action result for the BCI application.\\

\section{EXPERIMENTS}
To verify the feasibility of the proposed framework, we conducted experiments to evaluate the reliability of its two main tasks, user identification and intention classification, using private dataset.

\subsection{EEG Data Acquisition}
We collected EEG signals using three types of endogenous EEG paradigms (MI, SI, and VI) from eight subjects (four males and four females, aged 26.4 ± 1.7 years) to create a private dataset for our experiments. All subjects had no medical history of neurophysiological disorders and voluntarily participated in the experimental protocol, which was reviewed and approved by the Institutional Review Board of Korea University [KUIRB–2024–0065–01]. The subjects were instructed to perform three types of endogenous EEG paradigms while following instructions displayed on a monitor. Each experimental trial consisted of the following steps: \textit{i)} One of the target imagery classes (`apple', `star', `clover', or `snowman') and a target endogenous EEG paradigm (MI, SI, or VI) were presented on a monitor as an instruction for two seconds. \textit{ii)} A fixation cross appeared on the monitor for two seconds to stabilize the EEG response evoked by the instruction cue. \textit{iii)} A blank screen was displayed, and subjects imagined the target class using the instructed EEG paradigm for three seconds. \textit{iv)} Finally, a fixation cross was presented again for two seconds to eliminate any remaining EEG features evoked by the previous imagery task. Each subject performed 50 trials for each of the four imagery classes (200 trials in total) for all three EEG paradigms. During the experimental sessions, 32 EEG channels were placed on the subjects' scalps with conductive gel, following the 10/20 international placement system. The reference and ground electrodes were placed above the nasal bridge and on the Fpz channel, respectively. EEG signals were recorded using BrainVision software at a sampling rate of 250 Hz with a BrainAmp amplifier (Brain Products GmbH, Germany). The collected EEG signals were preprocessed using band-pass filtering to eliminate noise, and their amplitudes were rescaled by multiplying by $10^6$ to reduce the excessive influence of weight and bias parameters in the EEG decoding model.\\

\begin{table}[]
\small
\centering
\caption{Accuracy of User Identification by Endogenous Paradigms.}
\renewcommand{\arraystretch}{1.55}
\begin{tabular}{c c c c c}
\hline
\multirow{2}{*}{\textbf{Subject}} & \multicolumn{4}{c}{\textbf{Paradigm}}                               \\ \cline{2-5} 
                                  & \textbf{MI}    & \textbf{SI}    & \textbf{VI}    & \textbf{Overall} \\ \hline
S1                                & 0.999          & 0.998          & 0.999          & 0.999            \\ \hline
S2                                & 0.998          & 0.999          & 0.997          & 0.998            \\ \hline
S3                                & 0.998          & 0.994          & 0.996          & 0.996            \\ \hline
S4                                & 0.994          & 0.997          & 0.999          & 0.997            \\ \hline
S5                                & 0.993          & 0.992          & 0.994          & 0.996            \\ \hline
S6                                & 0.997          & 0.993          & 0.994          & 0.995            \\ \hline
S7                                & 0.991          & 0.998          & 0.991          & 0.993            \\ \hline
S8                                & 0.994          & 0.993          & 0.992          & 0.993            \\ \hline
\textbf{Mean}                     & \textbf{0.996} & \textbf{0.996} & \textbf{0.995} & \textbf{0.995}   \\ \hline

\end{tabular}
\end{table}

\subsection{EEG Decoding Models}
We used the ShallowConvNet \cite{schirrmeister2017deep} architecture as the EEG decoding model for both user identification and intention classification tasks, as it has demonstrated high versatility in decoding EEG features and is commonly used as a baseline model architecture in various tasks. For the user identification task, the model was trained to determine which subject the input EEG signals originated from. Similarly, for the intention classification task, the model was trained to classify which of the four imagery classes evoked the input EEG signals. Cross-entropy loss was used to optimize the model parameters for both tasks.\\

\subsection{Model Training}
To create the training dataset for the user identification task, we concatenated all EEG trials from all subjects and randomly permuted them. After permutation, the trials were split in a ratio of 70:10:20 for training, validation, and test sets, respectively. For the intention classification task, EEG trials from six subjects were used as the training set, while the remaining trials from two subjects were used as validation and test sets, respectively.
Both the user identification and intention classification models were trained with the same training parameters: a learning rate of 0.001, weight decay of 0.001, batch size of 64, using the Adam optimizer \cite{kingma2014adam} ($\beta_1=0.9$, $\beta_2=0.999$, and $\epsilon=10^{-8}$) for 100 epochs. During training, models were validated at every epoch using the validation set to prevent overfitting.\\

\subsection{Evaluation Method}
The model performance for each task was evaluated using average classification accuracy. For the user identification model, we evaluated performance by concatenating all EEG trials from all subjects and conducting five-fold cross-validation. Likewise, the intention classification model was also evaluated by using five-fold cross-validation for each subject, with a subject-dependent BCI setting.\\

\begin{table}[]
\small
\centering
\caption{Accuracy of Intention Classification by Endogenous Paradigms.}
\renewcommand{\arraystretch}{1.55}
\begin{tabular}{ccccc}
\hline
\multirow{2}{*}{\textbf{Subject}} & \multicolumn{4}{c}{\textbf{Paradigm}}                                                                                      \\ \cline{2-5} 
                                  & \multicolumn{1}{c}{\textbf{MI}}   & \multicolumn{1}{c}{\textbf{SI}}   & \multicolumn{1}{c}{\textbf{VI}} & \textbf{Overall} \\ \hline
S1                                & 0.40                              & 0.4                               & 0.35                            & 0.38             \\ \hline
S2                                & 0.47                              & 0.33                              & 0.35                            & 0.38             \\ \hline
S3                                & 0.45                              & 0.32                              & 0.34                            & 0.37             \\ \hline
S4                                & 0.95                              & 0.79                              & 0.37                            & 0.70             \\ \hline
S5                                & 0.51                              & 0.35                              & 0.37                            & 0.41             \\ \hline
S6                                & 0.33                              & 0.36                              & 0.37                            & 0.35             \\ \hline
S7                                & 0.95                              & 0.87                              & 0.59                            & 0.80             \\ \hline
S8                                & 0.38                              & 0.35                              & 0.38                            & 0.37             \\ \hline
\textbf{Mean}                     & \multicolumn{1}{c}{\textbf{0.56}} & \multicolumn{1}{c}{\textbf{0.47}} & \textbf{0.39}                   & \textbf{0.47}    \\ \hline
\end{tabular}
\end{table}

\section{RESULTS AND DISCUSSIONS}
\subsection{User Identification Performance}
Table I presents the experimental results for user identification tasks. As indicated in the table, the overall performance of user identification was highly reliable across all participants and all endogenous EEG paradigms, achieving an average classification accuracy of 0.995. These findings suggest that EEG signals can be utilized as a robust modality for identifying user-specific information and preferences, based on distinctive EEG features unique to each individual.\\

\subsection{Intention Classification Performance}
Table II presents the results of the experiment for the intention classification task. The overall classification accuracy across all paradigms was 0.47. Among the three endogenous paradigms, MI demonstrated the most reliable performance compared to SI and VI, with a mean performance of 0.56, despite using a simple ShallowConvNet model as the decoding model without any task-specific methods or fine-tuning. For SI, the mean performance was 0.47, while VI showed the lowest performance with a mean accuracy of 0.39. These results suggest that MI and SI could potentially serve as reliable endogenous paradigms for representing user intention as an intuitive communication method.
Among the participants, subject 6 exhibited the highest performance across all three paradigms (MI: 0.95, SI: 0.87, and VI: 0.59) compared to other subjects. Subject 4 also achieved reliable performance in MI and SI (0.95 and 0.79, respectively), although the performance in VI was lower. Other subjects (1, 2, 3, 5, 7, and 8) did not achieve as high performances as subjects 6 and 4, but still demonstrated the potential for using endogenous EEG paradigms, with accuracies above chance levels.\\

\section{CONCLUSION}
In this paper, we proposed a conceptual framework for a personalized BCI application, which can identify users and provide services based on user information, using endogenous paradigms. Also, we demonstrated the feasibility of the two main tasks for our framework, user identification, and intention classification, with the experiments using private datasets. For the further enhancement of personalized services, incorporating user feedback mechanisms using additional biosignals, such as EMG, along with historical user activity data, has the potential to refine recommendation algorithms and provide more precise and reliable personalized services. Our future research will focus on improving classification performance through more advanced EEG decoding models, incorporating task-specific optimizations, and implementing the proposed framework as a tangible real-time BCI application with integrating user feedback mechanisms and leveraging user database information to enhance personalization \\
\bibliographystyle{IEEEtran}
\bibliography{MANUSCRIPT}

\end{document}